\documentclass[aps,twocolumn,superscriptaddress,nofootinbib]{revtex4}

\usepackage{amsfonts}
\usepackage{amssymb}
\usepackage{mathdots}

\usepackage{graphicx}
\usepackage{bbm}
\usepackage{color}
\usepackage{pifont}
\usepackage{enumerate}
\usepackage{subfigure}
\usepackage{color}
\newcommand*{\rom}[1]{\expandafter\@slowromancap\romannumeral #1@}
\makeatother
\usepackage{mathptmx}
\usepackage{mathrsfs}
\usepackage{mathtools}
\usepackage{amsmath}
\usepackage{amsfonts}
\usepackage{amssymb}
\usepackage{mathdots}
\usepackage{bbold}
\usepackage{isomath}
\usepackage{mathrsfs}
\usepackage{amsthm} 
\usepackage{graphicx}
\usepackage{bbm}
\usepackage{color}
\usepackage{pifont}
\usepackage{enumerate}
\usepackage{amsthm}
\theoremstyle{plain}
\theoremstyle{definition}
\usepackage{mathtools}
\makeatletter
\newsavebox{\@brx}
\newcommand{\llangle}[1][]{\savebox{\@brx}{\(\m@th{#1\langle}\)}%
  \mathopen{\copy\@brx\kern-0.5\wd\@brx\usebox{\@brx}}}
\newcommand{\rrangle}[1][]{\savebox{\@brx}{\(\m@th{#1\rangle}\)}%
  \mathclose{\copy\@brx\kern-0.5\wd\@brx\usebox{\@brx}}}
\makeatother
\usepackage[titletoc]{appendix}
\usepackage[english]{babel}
\usepackage[autostyle]{csquotes}
\usepackage{hyperref} 
\usepackage{cleveref} 

\begin{document}

\title{Implementing a Noise Protected Logical Qubit in Methyl Groups via Microwave Irradiation}


\author{R. Annabestani}
\email[]{rannabes@uwaterloo.ca}
\affiliation{Institute for Quantum Computing, University of Waterloo, Waterloo, Ontario N2L 3G1, Canada}
\affiliation{Department of Physics and Astronomy, University of Waterloo, Waterloo, Ontario N2L 3G1, Canada}
\author{D. G. Cory}
\email[]{dcory@uwaterloo.ca}
\affiliation{Institute for Quantum Computing, University of Waterloo, Waterloo, Ontario N2L 3G1, Canada}
\affiliation{Department of Physics and Astronomy, University of Waterloo, Waterloo, Ontario N2L 3G1, Canada}
\affiliation{Department of Chemistry, University of Waterloo, Waterloo, Ontario N2L 3G1, Canada}
\affiliation{Perimeter Institute for Theoretical Physics, Waterloo, Ontario N2L 2Y5, Canada}
\affiliation{Canadian Institute for Advanced Research, Toronto, Ontario  M5G 1Z8, Canada}



\begin{abstract}
We propose a proof-of-principle experiment to encode one logical qubit in noise protected subspace of three identical spins in a methyl group. The symmetry analysis of the wavefunction shows that this fermionic system exhibits a symmetry correlation between the spatial degree of freedom and the spin degree of freedom. We show that one can use this correlation to populate the noiseless subsystem by relying on the interaction between the electric dipole moment of the methyl group with a circularly polarized microwave field. Logical gates are implemented by controlling both the intensity and phase of the applied field.
 \end{abstract}


\maketitle
\section{Introduction}
A quantum bit, known as qubit, is a two-level coherent system \cite{NCu00, Sch95}. A qubit can store quantum information for use in quantum computing, quantum communication, sensing and etc \cite{NCu00}. Over the past decades, there have been proposals for physical realization of a logical qubit including cold atoms, polarization of a single photon and spins. What is common between all of these physical systems is that there always exist some undesired couplings with their environment, which manifest  themselves  as a noise process on the system of interest. These undesired couplings shorten both relaxation time and coherence time of a qubit that is equivalent to losing quantum information. \\

One  of  the  approaches  for  protecting  the  quantum  information is to encode it in such a way that the logical qubit is not affected by noise in the first place. In fact, the notion of decoherence free subspace or noiseless subsystems have been developed, where the information is encoded in a particular part of the Hilbert space so as to have noise immunity \cite{VKL01, L14, LW03}. For example, when the logical states of a qubit are considered as the actual physical states of a spin half particle, one should expect both X-noise (spin flip) and Z-noise (dephasing) which corrupt the information, \cite{NCu00}. But, if one considers a group of spins, there are other degrees of freedom rather than the  spin magnetization that can be used as a logical basis that is protected against majority of noise processes \cite{V01}. \\
  
Nuclear Magnetic Resonance (NMR), owing to its long history, has been  among the  first  physical  candidates  to  experimentally demonstrate the quantum information and error correction algorithms \cite{VSBYC01, WLC01, SchV98, C00}. We investigate an NMR implementation of a noiseless logical qubit that is created by using the collective properties of a group of three indistinguishable spins. In particular, we consider a methyl group, which consists of a carbon symmetrically bonded to three protons, and show that the group of protons can serve as one logical qubit that is protected against collective noise.\\

 Since the early age of NMR, the molecular motion and the spin properties of methyl groups have been studied both theoretically and experimentally \cite{F65, AH70, C71, H99}. Methyl groups have been used for experimental demonstration of classical states with long relaxation times that are known as long lived states (LLS) \cite{IB12, M13}. In this report, we show that LLS can be used as an initial step for implementing a coherent superposition of logical basis states. In particular, we analyze the symmetry of the internal rotation and the spin Hamiltonian of methyl groups using a more modern mathematical language which may benefit both the NMR community and the quantum information community. We then use that symmetry analysis to proposes a practical way of converting LLS into a logical qubit that can be used for storing quantum information and is immune to collective noise. \\

There is a challenge for accessing the noise protected logical basis of methyl groups. We show that the noise immunity is due to the spin symmetry of indistinguishable spins and the indistinguishability of spins results in the degeneracy of the subspace of interest. As a result, the logical states are not individually accessible. Nevertheless, we take advantage of the molecular symmetry and the microwave spectroscopy technique \cite{TS75, JAM03, BW03} to propose an experiment that leads to populating and controlling the noise protected subspace of methyl groups without breaking the degeneracy. \\

In particular, we show that there is a symmetry correlation between the rotational degree of freedom and the spin degree of freedom due to Pauli exclusion principle.  This space-spin correlation suggests that one can populate the noiseless logical states, albeit directly inaccessible due to their degeneracy, by controlling the other degrees of freedom such as rotational motion. We use a symmetry argument based on the group theory to derive the allowed transitions under an electric field irradiation and conclude that an interaction with a \textit{right} or \textit{left} circularly polarized microwave filed results in populating the logical \textit{zero} or \textit{one} states individually without lifting their degeneracy. Moreover, we show a delicate control of the intensity and phase of the applied field can lead to implementing the logical gates. Therefore, we demonstrate a proof-of-principle experiment to use methyl groups for implementing one logical qubit that has a long relaxation and coherence times. \\

In section.\ref{Sec_Logical_Qubit}, we define the logical qubit for methyl groups by looking at the permutation symmetry and the spin Hamiltonian of three indistinguishable spins. Then, in section.\ref{Sec_Int_Rot}, we study the internal rotational motion of $CH_{3}$, also known as a \textit{rotor}, followed by section.\ref{Sec_Tot_Sym} that argues the symmetry of the total wavefunction. Given this symmetry analysis, section.\ref{Sec_protected_state_Methyl} seeks for novel means of accessing and controlling the noise protected subspace using the microwave irradiation, leading to preparation and manipulation of one logical qubit. 

\section{Logical Qubit in Methyl Groups}
\label{Sec_Logical_Qubit}
\subsection{Three Indistinguishable Spins}
Three indistinguishable spins form a $\textbf{C}_{3}$ group because they are invariant under the cyclic permutation operator, which is defined by
 \begin{eqnarray}
  \hat{\mathit{P}}_{+} | i\rangle \otimes | j\rangle  \otimes | k \rangle &=&  | k\rangle  \otimes | i\rangle  \otimes | j \rangle,\\\nonumber
  \hat{\mathit{P}}_{-} | i\rangle  \otimes | j\rangle \otimes | k \rangle &=&  | j\rangle  \otimes | k\rangle  \otimes | i \rangle,
 \end{eqnarray}
 for $\forall  | i\rangle, |j\rangle , |k\rangle \in\{ |\uparrow\rangle, |\downarrow\rangle\}$. It is straightforward to find a matrix representation of $\hat{\mathit{P}}_{\pm}$  and expand it in Pauli operator basis,
\begin{eqnarray}
\hat{\mathit{P}}_{\pm}= \frac{ 1}{4} \Big( \mathbb{1}&+& \vec{\sigma}^{(1)}. \vec{\sigma}^{(2)} + \vec{\sigma}^{(2)}. \vec{\sigma}^{(3)}  + \vec{\sigma}^{(1)}. \vec{\sigma}^{(3)} \\ \nonumber
&\mp & \textit{i} \sum\limits_{\alpha\beta\gamma} \epsilon_{\alpha\beta\gamma} \  \sigma_{\alpha}  \otimes  \sigma_{\beta} \otimes  \sigma_{\gamma}\Big),
\end{eqnarray}
 where $\epsilon_{\alpha\beta\gamma}$ is the Levi-Civita coefficient with $\alpha, \beta, \gamma \in \{ x, y, z\}$ and $\sigma$s are Pauli matrices. The complex eigenvalues of the $\hat{\mathit{P}}_{\pm}$ are $\{ 1, \varepsilon , \varepsilon ^{*} \}$  with $\varepsilon= e^{2\pi i /3}$ that correspond to three irreducible representations $\{A, E_{+}, E_{-}\}$ respectively. We represent the eigenstates of $\hat{\mathit{P}}_{+}$ with $| s,m\rangle$ where the $s$ indicates the symmetry irreducible representations, $s \in \{ A, E_{+}, E_{-} \}$ and the $m$ is the eigenvalue of the z component of the total spin angular momentum operator, $\hbar \hat{\textbf{S}}_{z}=\frac{\hbar}{2} \sum\limits_{i=1}^{3} \sigma ^{(i)}_{z}$. We name $\{ |s, m\rangle \}$ the Cyclic Permutation (CP) basis whose explicit expansion in terms of the computational basis is given in Table.\ref{Tab_3spins_symmetry}.\\

 \begin{table}[h]
 \centering
 \begin{tabular}{c|c|c}
 $ | s, m\rangle $  & Expansion in $\{ |\uparrow\rangle, |\downarrow\rangle\} ^{\otimes 3}$ & $\hat{\mathit{P}}_{+}$'s Eigenvalue  \\
 \hline
 $|A, 3/2\rangle $ &    $|\uparrow \uparrow  \uparrow\rangle$ & 1 \\
$|A, 1/2\rangle $ &  $\frac{1}{\sqrt{3}}\left(|\uparrow \uparrow  \downarrow\rangle+ |\downarrow \uparrow \uparrow  \rangle+ |\uparrow  \downarrow \uparrow \rangle\right)$& 1 \\
$|A, -1/2\rangle $ &  $\frac{1}{\sqrt{3}}\left(|\downarrow \downarrow  \uparrow\rangle+ |\uparrow \downarrow \downarrow  \rangle+ |\downarrow  \uparrow \downarrow \rangle\right)$& 1 \\
$|A, -3/2\rangle $ & $| \downarrow \downarrow \downarrow\rangle$& 1  \\
  \hline
$|E_{+}, 1/2\rangle $ & $\frac{1}{\sqrt{3}}\left(|\uparrow \uparrow  \downarrow\rangle+  \varepsilon^{*}|\downarrow \uparrow \uparrow  \rangle+\varepsilon |\uparrow  \downarrow \uparrow \rangle\right)$ &$\varepsilon$ \\
$|E_{+}, -1/2\rangle $ &  $\frac{1}{\sqrt{3}}\left(|\downarrow \downarrow  \uparrow\rangle+ \varepsilon^{*} |\uparrow \downarrow \downarrow  \rangle+\varepsilon |\downarrow  \uparrow \downarrow \rangle\right)$ &$\varepsilon$ \\
  \hline
$|E_{-}, 1/2\rangle $ & $\frac{1}{\sqrt{3}}\left(|\uparrow \uparrow  \downarrow\rangle+ \varepsilon|\downarrow \uparrow \uparrow  \rangle+\varepsilon^{*} |\uparrow  \downarrow \uparrow \rangle\right)$ &$\varepsilon^{*}$\\
$|E_{-}, -1/2\rangle $ & $\frac{1}{\sqrt{3}}\left(|\downarrow \downarrow  \uparrow\rangle+ \varepsilon |\uparrow \downarrow \downarrow  \rangle+ \varepsilon^{*} |\downarrow  \uparrow \downarrow \rangle\right)$ & $\varepsilon^{*}$ \\
  \end{tabular}
  \caption{The cyclic permutation basis expanded in the computational basis. $\hbar=1$ is considered.} \label{Tab_3spins_symmetry}
 \end{table}
 
The Hilbert space of these $3$ spins is decomposed to two subspaces, $\mathcal{H}= \bigoplus\limits_{j} \mathcal{H}_{j} = \mathcal{H}_{3/2} \oplus \mathcal{H}_{1/2}$, where $\text{dim}(\mathcal{H}_{3/2})=4$ and $\text{dim}(\mathcal{H}_{1/2})=4$. Each subspace $\mathcal{H}_{j}$ is further decomposed into a product of two \textit{subsystems}, where the first component refers to the $s$ label and the second component refers to the $m$ label,
\begin{equation}
 \mathcal{H}= \bigoplus\limits_{j} \mathbb{C}_{\textbf{s}_{j}} \otimes \mathbb{C}_{\textbf{m}_{j}}= \mathbb{C}_{1} \otimes \mathbb{C}_{4} \oplus \mathbb{C}_{2} \otimes \mathbb{C}_{2}.
\end{equation}
  We used the bold notation to distinguish the dimension of the space of a label from its value, i.e., $\textbf{s}_{j} = \text{dim}(\{s_{j}\})$ and  $\textbf{m}_{j} = \text{dim}(\{m_{j}\})$. 
Any collective noise operator that does not distinguish spins acts trivially on the symmetry label. More precisely, all components of the total spin angular momentum, $ \hat{\textbf{S}}_{ \alpha}$ with $\alpha \in \{ x, y, z\}$, have a block-diagonal form in the CP basis,
\begin{eqnarray}
\label{Eq_symmetry-expansion}
 \hat{\textbf{S}}_{\alpha}&=& \sum\limits_{m,m'=-\frac{3}{2}}^{\frac{3}{2}}   a_{mm'} \ |A , m \rangle \langle A, m' |\\ \nonumber  &+& \sum\limits_{m,m'=-\frac{1}{2}}^{\frac{1}{2}}  b_{mm'} \left(\ |E_{+} , m \rangle \langle E_{+}, m' | + \ |E_{-} , m \rangle \langle E_{-}, m' |\right).
\end{eqnarray}
Being block diagonal shows that the symmetry label is \textit{preserved} by the collective spin operators. Therefore, one can choose the \textit{symmetry label} subsystem of the $E$ subspace ( or $ j=1/2$) of 3 identical physical qubits to encode one logical qubit because it is \textit{protected} against collective noise. Explicitly, the logical basis is defined by
\begin{eqnarray}
\label{Eq_logic-basis}
|\bar{0}\rangle &\equiv & | E_{+} \rangle \otimes |\phi_1\rangle,\\ \nonumber
|\bar{1}\rangle &\equiv & |E_{-} \rangle \otimes |\phi_2 \rangle,
\end{eqnarray}
where $|\phi_{1(2)}\rangle$ is an arbitrary pure state expanded in the $\mathbb{C}_{\mathbf{m}_{1/2}}= \text{Span}\{|\pm 1/2\rangle\}$ subsystem. Given an arbitrary pure state $|\psi\rangle = a | \bar{0}\rangle + b |\bar{1}\rangle$ with $\sqrt{|\alpha|^2+ |\beta|^2}=1$, the logical qubit is obtained by a partial trace over the $\mathbb{C}_{\textbf{m}_{1/2}}$ subsystem as
  \begin{eqnarray}
  \label{Eq_partial-noisy}
  \rho_{\text{logic}} = Tr_{m}[ |\psi \rangle \langle \psi |] =\left(\begin{array}{cc}
    |a|^2  &  a^{*}b\langle\phi_{1}| \phi_{2}\rangle\\
   b^{*}a \langle\phi_{2}| \phi_{1}\rangle & |b|^2
  \end{array}\right).
  \end{eqnarray}
Note that the off-diagonal terms are proportional to the overlap of $|\phi_{1}\rangle$ and $|\phi_{2}\rangle$. This implies that if we prepare $|\phi_{1}\rangle = |\phi_{2}\rangle $ perfectly, $\rho_{\text{logic}}$ implements an ideal logical qubit, otherwise, any imperfection in the preparation acts as a decoherence process. Note that this type of decoherence is purely due to the imperfections in the state preparation step, not the external noise, because all $\hat{\textbf{S}}_{\alpha}$ with $\alpha \in \{x,y,z\}$ preserve $\langle\phi_{2}| \phi_{1}\rangle$.

    \subsection{Spin Hamiltonian of a Methyl Group}
  \label{Sec_LLS}
 In the presence of a magnetic field and in the absence of any chemical shift anisotropy (CSA) and/or any dipole-dipole interaction (DD), the spin Hamiltonian of the three protons in a methyl group is
 \begin{equation}
 \label{Eq_Spin_Ham}
 H_{\text{spin}}= \frac{\omega_{h}}{2} \sum\limits_{i=1}^{3} \ \sigma^{(i)}_{z} + 2\pi\ J_{0}\ \sum\limits_{j< k} \  \vec{\sigma}^{(j)}.\vec{\sigma}^{(k)},
 \end{equation}
 where $\omega_{h}= \gamma_{h}B_{0}$ is the proton frequency and $J_{0}$ is the scalar coupling constant between any two protons, which is normally small compared to the Zeeman energies. The spin Hamiltonian in Eq.\ref{Eq_Spin_Ham}, does not distinguish these identical protons, and thus, $[H_{\text{spin}}, \hat{\mathit{P}}_{\pm}]=0$. This commutation relation implies that spin Hamiltonian preserves the symmetry label, $s$. Therefore, at relatively high field, where the DD couplings and/or the CSA are negligible and for a noise model that the methyl group interacts symmetrically with the environment, neither the environment nor the local spin Hamiltonian in Eq.\ref{Eq_Spin_Ham} breaks the cyclic permutation symmetry. Thus, a methyl groups is a candidate for storing one logical qubit that exhibits long relaxation and decoherence times because it is immune to all collective noise.\\
 
However, there is a delicate point here that prevents us from doing so. In one hand, the protection against collective noise has its roots in the indistinguishably of spins which allows us to consider the symmetry label $E_{+}/E_{-}$ as a noiseless subsystem. On the other hand, if neither the the local spin Hamiltonian in Eq.\ref{Eq_Spin_Ham} nor the coupling to the environment distinguishes spins, the spin eigenstates $|E_{+},m \rangle$ and $|E_{-},m \rangle$ are degenerate, and thus, the logical basis $|\bar{0} \rangle$ and $|\bar{1} \rangle$ are not accessible individually. When the logical basis states are not individually accessible, implementing any arbitrary logical qubit is not feasible. Nevertheless, in the following we show that it is possible to populate these degenerate states individually without breaking the cyclic permutation symmetry. We propose to use other degrees of freedom of methyl groups, such as internal rotation, to indirectly access and control the degenerate subspace of the spin degree of freedom.  Thus, in the next section, we review the symmetry of the rotational degree of freedom of a methyl group.

\section{Internal Rotation of Methyl Groups}
\label{Sec_Int_Rot}
 This section considers the internal rotation or the \textit{torsional} degree of freedom of a methyl group which has been extensively studied in literature, \cite{F65, AH70, C71, H99}. Idealistically, a methyl group is a \textit{free rotor} that is a rigid body \textit{freely} rotating around the $z$ axis and its Hamiltonian is simply the $z$ component of the total angular momentum,
\begin{equation}
\label{Eq_Torsion_Ham}
H_{\text{rot}}= \hat{L}_{z}= \frac{- \hbar^2}{2 I_{0}} \frac{\partial^2}{\partial \varphi^{2}},
\end{equation}
where $I_{0} $ is the moment of inertia and the angle $\varphi$ is conventionally defined as the azimuthal angle between a proton and a reference axis in the molecular framework. The eigenfunctions of this free rotor are  $ \frac{e^{ \pm i l \varphi}}{\sqrt{2}} $ with corresponding eigenenergies $E_{l} = \frac{\hbar^2}{2 I_{0}}\  l^{2}$ where $l \in \{ 0, \pm 1, \pm 2, \dots\}$. The constant $F= \frac{\hbar^2}{2 I_{0}}$ is  referred as the \textit{free rotor energy} constant.\\

\begin{figure}[h]
  \centering
\includegraphics[width=0.5\linewidth]{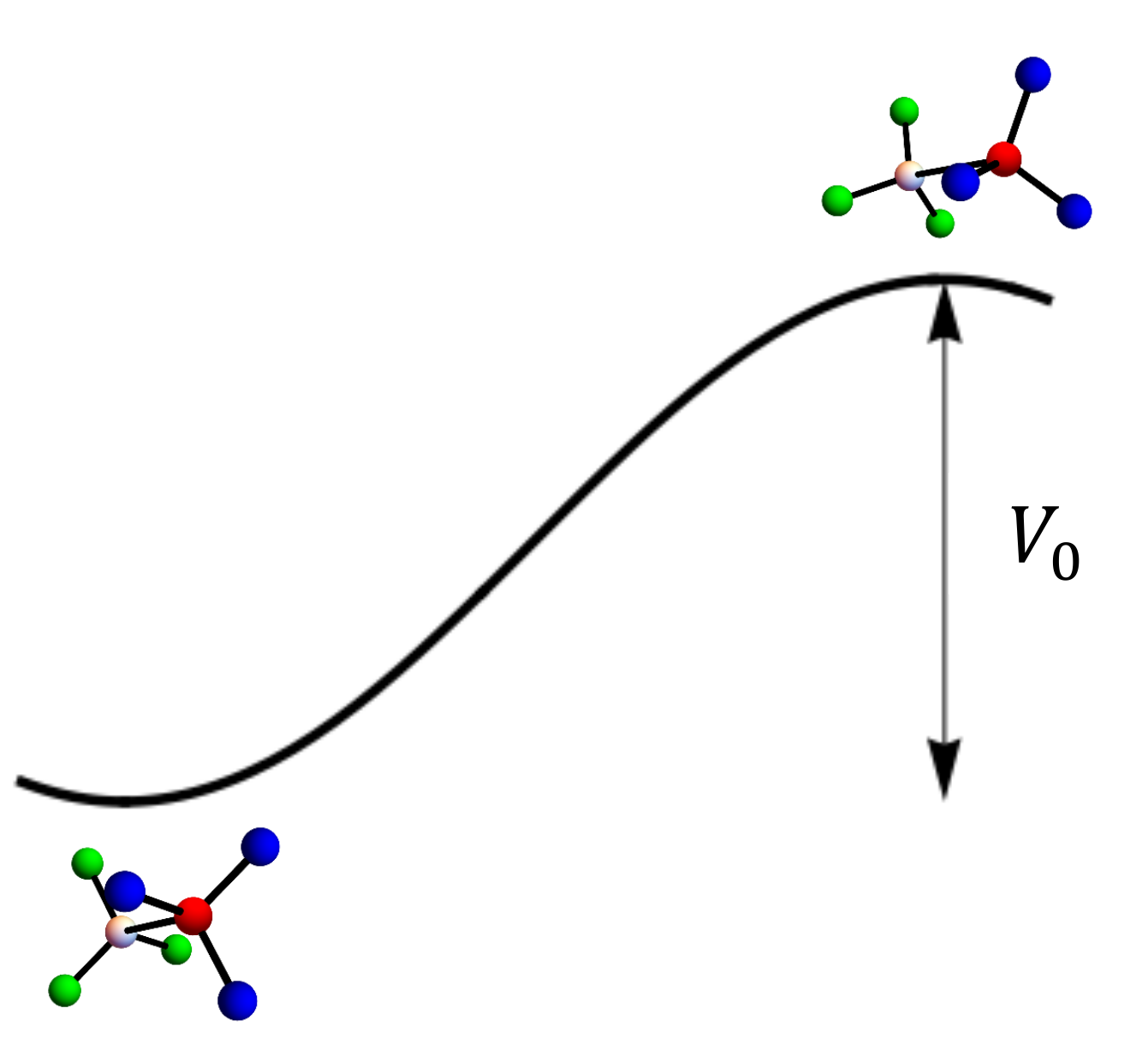}
  \caption[Ethane Hindering Potential]{Ethane Hindering Potential: The barrier height is the amount of energy that is required to move from the staggered configuration to the eclipsed configuration. }
  \label{fig_Potential_Ethane}
\end{figure}

Realistically, a methyl group in a symmetric top molecules like $X-CH_{3}$ does not rotate freely due to the existance of a \textit{hindering} potential which is imposed by the rest of the molecule as well as other external molecules\footnote{ The main source of hindering potential depends on the molecule. For example, in case of ethane van der Waals interactions and hyperconjugation have been reported in literatures \cite{P83} and \cite{PG01}.}. In ethane, for example, the molecule prefers to be in the staggered configuration rather than the eclipsed configuration, \cite{PG01}. The amount of energy that is required to move from one configuration to another defines the height of the hindering potential or the \textit{barrier} (Fig.\ref{fig_Potential_Ethane}). Therefore, in addition to the free rotation term, there is an extra potential that affects the rotational motion \cite{H99},   
\begin{equation}
H_{\text{tor}}= \frac{- \hbar^2}{2 I_{0}} \frac{\partial^2}{\partial \varphi^{2}} + V_{\text{h}}(\varphi) ,
\end{equation}
in which
\begin{equation}
V_{\text{h}}(\varphi)=\sum\limits_{k}\ \frac{V_{3k}}{2} \left(1- \cos[ 3k\ \varphi]\right), \hskip 0.5 cm \text{with} \hskip 0.5 cm  k= 1, 2, \dots
\end{equation} 

 Depending on the geometry of the molecule, the hindering potential has $3$ fold symmetry, $6$ fold symmetry and etc. But, often the first term $V_{3}=V_{0}$ is the dominant one and all other $V_{3k\neq 3}$ are negligible. For the following discussion we consider a 3-fold potential only.\\
 
  In the \textit{firm rotor} limit or a rigid rotor with extremely high rotational barrier, the motion along the $\varphi$ direction is very restricted, because, the barrier height is much larger than the free rotor energy,  i.e., $ V_{0}\gg F$. Thus, we approximate $V(\varphi)$ with three quantum wells and treat each well as a harmonic potential, 
 \begin{equation}
V_{\text{h}} \approxeq \frac{V_{0}}{2} \left(1- \cos[ 3\ \varphi]\right) \approxeq \frac{ V_{0}}{2}\left( \frac{(3\varphi)^{2}}{2!} - \frac{(3\varphi)^{4}}{4!} +\dots \right).
 \end{equation}
Reasonably, each level has the degeneracy of 3, because there are three wells, and so, three harmonic oscillators. The corresponding 3-fold degenerate eigenenergies are $E_{n}\approxeq 3\sqrt{\frac{V_{0}}{I_{0}}} (n + \frac{1}{2})$ with $n=0, 1, 2, \dots.$  \\

\begin{figure}[h]
  \centering
\includegraphics[width=1\linewidth]{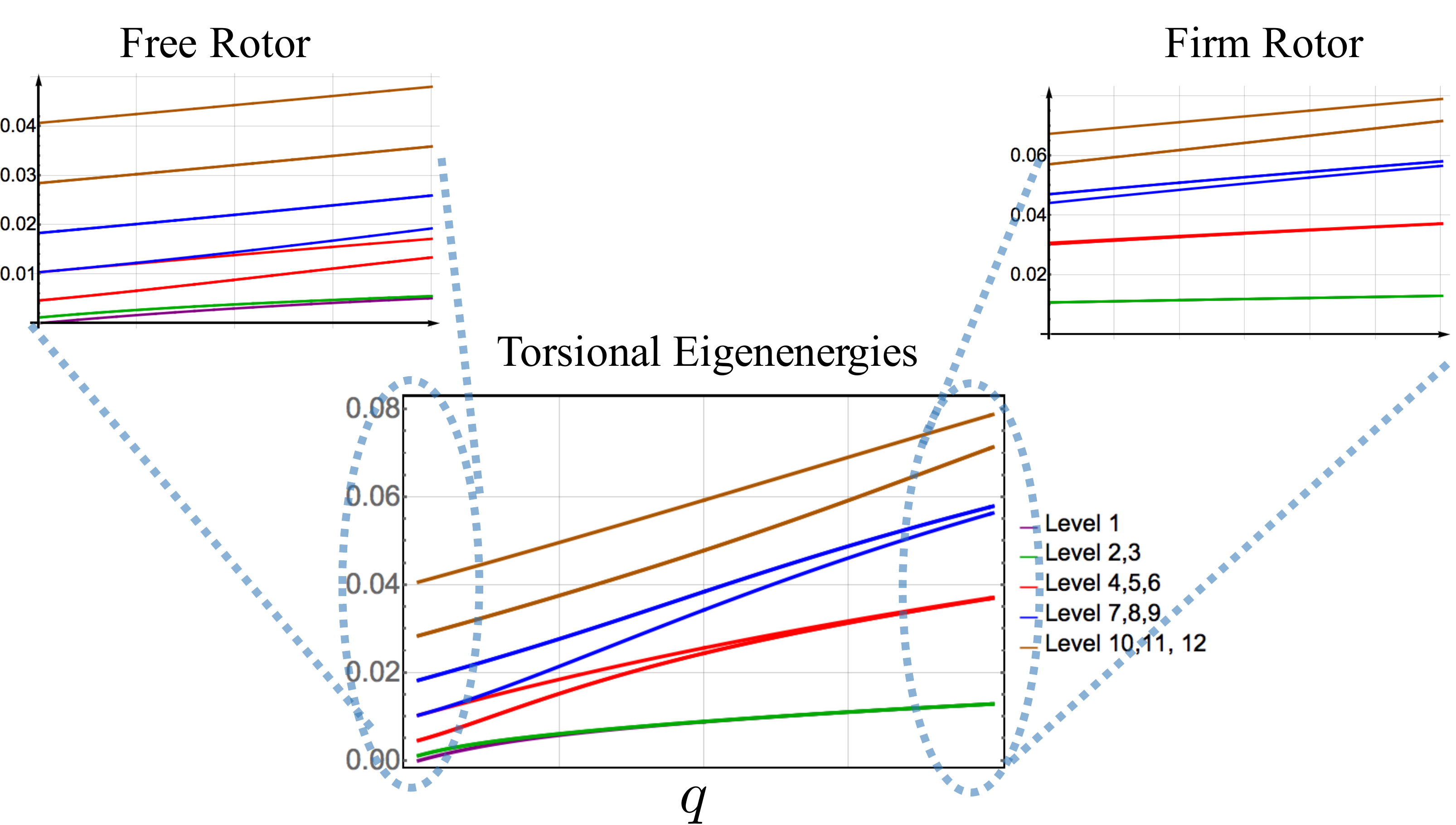}
  \caption[Torsional Eigenenergies of a methyl group]{ Torsional Eigenenergies of a Methyl Group: The left side, $q=0$, is the free rotor limit when the hindering potential is very shallow or nearly zero and the right side, $q=1$, is the firm rotor when the barrier height is very large compared to the free rotor energy.}
  \label{fig_rotor}
\end{figure}

In non-extreme cases, where the methyl group is neither a free rotor nor a firm rotor (harmonic oscillator), we solve the Sch\"{o}dinger equation numerically and plot the eigenenergies as a function of a parameter $0< q < 1$, where the $q=0$ corresponds to a free rotor ($V_{0}=0$) and the $q=1$ corresponds to a firm rotor ($V_{0} \gg F$). The first twelve eigenenergies of a methyl group are demonstrated in Fig.\ref{fig_rotor}, from the free rotor limit to the firm rotor limit. The left plot is associated with the nearly free rotor ($V_{0}\approx0$), and as expected, other than the ground state ($l=0$), all other energy levels are double degenerate. At an intermediate regime, when $ V_{0}\geq F$, every three energy levels tend to group together and form a \textit{band} that consists of a non-degenerate level and a double degenerate level. Thus, at $ V_{0}\geq F$ limit we label the internal rotation eigenstates with $\Phi_{\lambda, n}$ with corresponding energies $E_{\lambda, n}$ in which $n$ refers to the band level (harmonic oscillator level) and $\lambda \in \{0, \pm1 \}$. At each band $n$, we denote the \textit{internal} energy difference between the $\lambda=0$ level and the $\lambda=\pm1$ levels with $\Delta E_{n}$. At a particular value of $q$, we observe that the sign of the energy \textit{splitting}, $\Delta E_{n}$, changes from the $n^{\text{th}}$ band to the $(n+1)^{\text{th}}$ band, and its amplitude gets larger and larger as we go higher in $n$. Because of these two observations, it is reasonable to see at some point the $E_{\lambda =0,n}$ level and $E_{\lambda =0, n+1}$  merge together and form a double degenerate level. Indeed, even in the intermediate regime where $V_{0}\geq F$, the high energy levels behave like that of a free rotor and are doubly degenerate. Because in that energy scale, the methyl group effectively \textit{does not see} the barrier. These numerical observations are better justified in the upcoming section. For larger $q$ (or larger barrier height), the internal splitting $\Delta E_{n}$ gets smaller and smaller and eventually it becomes zero at the firm rotor limit, yielding to a 3-fold degenerate level. The right plot of Fig.\ref{fig_rotor} is associated with the nearly firm rotor ($V_{0}\gg F$). 

Each molecule has a certain geometry and mass, and so, has a certain free rotation energy, $F$. It also has an associated hindering potential with specific barrier height, $V_{0}$. Thus, depending on the ratio $\frac{V_{0}}{F}$, an organic compound may be considered as a free rotor, or an intermediate rotor, or an firm rotor based on the structure of the lowest energy levels. Table.\ref{Tab_rotor} lists some examples of chemical compounds with their corresponding barrier heights, free rotor energy constant and the splitting of the ground state $n=0$. 
\begin{table}[h]
\begin{center}
\begin{tabular}{c|c|c|c|c|c}
Compound & $V_{0}(meV)$  & $F(meV)$  & $\Delta E_{0} (GHz)$ & Ref & Regime\\
\hline
o-fluorotoluene & 28.14  & 0.64&  20 &\cite{JAM03} & Intermediate\\ 
m-fluorotoluene & 2.6  & 0.66 &  860 &\cite{RTN68} & Free Rotor\\ 
Lithium acetate &  & & 60 & \cite{IB12} & \\
4-Methylpyridine& 7.3 & & 120 &\cite{PH97} & Intermediate\\
o-toluidine &86.79 & 0.66 & 350 &  \cite{OMI87} , \cite{KVSM09} & Intermediate\\
\end{tabular}
\end{center} 
\caption[Examples of chemical compounds with their corresponding barrier height and free rotor energy]{Examples of chemical compounds with their corresponding barrier height and free rotor energy. In the third column, the $\Delta E_{0}$  is the energy splitting between $(\lambda=\pm1,n=0)$ and $(\lambda =0,n=0)$ levels of the torsional states (or the splitting between $l=0$ and $l=\pm$ levels of the rotational states, in the limit of the free rotor).} 
 \label{Tab_rotor}
\end{table}
\subsection{The Symmetry of the Torsional States}
 A methyl group forms a $\textbf{C}_{3}$ group in the spatial space that has three irreducible representations, $\{A, E_{+}, E_{-}\}$, with their corresponding character values $\{1, \epsilon,  \epsilon^{*}\}$ with $\epsilon=e^{ i\frac{2\pi}{3}}$. In previous section, we solved the Schr\"{o}dinger equation numerically, yielding the torsional eigenfunctions  $\Phi_{ \lambda,n} (\varphi)$ in the intermediate rigid rotor limit ($ V_{0}\geq F$). Here, we analyze the symmetry of $\Phi_{ \lambda,n} (\varphi)$, and show that the label $\lambda$ is associated to one of the irreducible representations of $\textbf{C}_{3}$ group. \\
 
\begin{figure}[h]
  \centering
\includegraphics[width=1\linewidth]{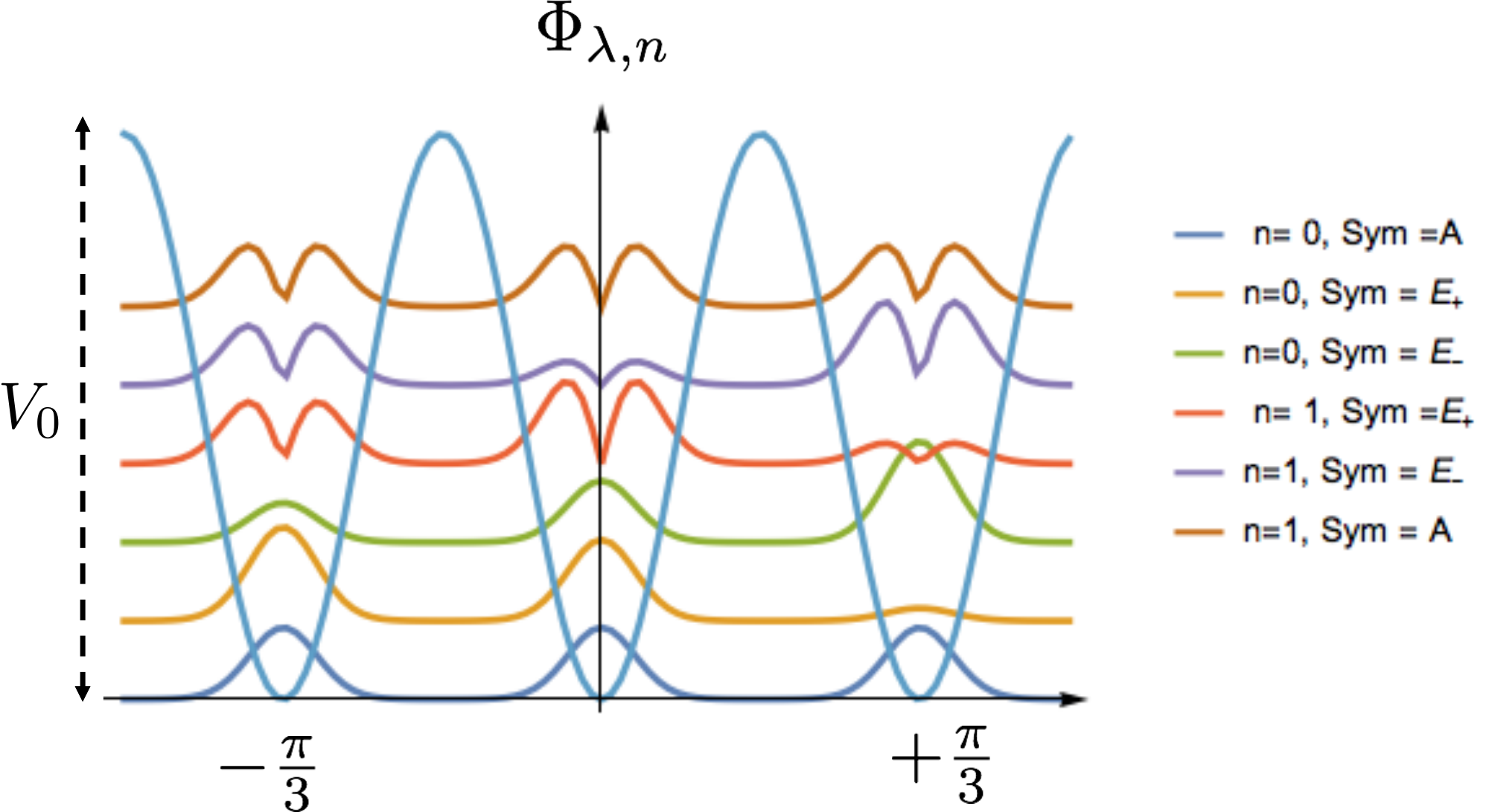}
  \caption[Torsional Eigenfunctions]{  \label{fig_ES_Methyl_Group}Torsional Eigenfunctions: The first 6 eigenfunctions of the torsional Hamiltonian are presented. The first and sixth levels are totally symmetric ($A$) and the rest have $E_{\pm}$ symmetry.}
\end{figure}

 We expand $\Phi_{ \lambda,n}$ in terms of free rotor eigenfunctions which form an orthonormal basis, 
 \begin{equation}
   \Phi_{\lambda,n} (\varphi)= \sum\limits_{l} c_{l} \ \frac{e^{ i l \varphi}}{\sqrt{2}}.
 \end{equation}
If an eigenfunction $\Phi$ has \textit{$A$ symmetry}, it must be invariant under the $\frac{2\pi}{3}$ rotation. Therefore, the non-zero terms in the above expansion must be $l = 3\ k $, where $k$ is an integer. Similarly, if it has \textit{$E_{\pm}$ symmetry}, under the $\frac{2\pi}{3}$ rotation it must pick up a phase $e^{\pm i\frac{2\pi}{3}}$, i.e., $ \Phi_{\lambda,n} (\varphi\pm \frac{2\pi}{3})= e^{\pm i\frac{2\pi}{3}}\    \Phi_{\lambda,n} (\varphi)$. Therefore, the non-zero terms in the above expansion must be $l =3\ k +1$ for $E_{+}$ symmetry and $ l = 3\ k -1$ for $E_{-}$ symmetry. This property can be used as a test to see which kind of symmetry each numerical eigenfunction has, $A$ or $E_{\pm}$ or a combination. We realize that for a methyl group with an intermediate barrier, at the $n^{\text{th}}$ band, the $\lambda=0$ state has only $A$ symmetry and the the doubly degenerate states, $\lambda=\pm1$, have $E_{\pm}$ symmetry. Because $E_{\pm}$ are degenerate, any superposition of them is also a solution to the Schr\"{o}dinger equation. The first 6 numerical eigenfunctions of the torsional Hamiltonian are plotted in Fig.\ref{fig_ES_Methyl_Group} for the choice of $V_{0} =32$ meV and $F= 0.6$ meV.\\

An alternative approach for finding the torsional eigenfunctions is to treat $V_{\text{h}}(\varphi)$ as a periodic lattice with three wells. According to the Bloch theorem \cite{B28}, given a periodic potential $V(x)$ with $N$ wells (atoms), the lattice (or the molecule) wavefunction, $\Phi(x)$, has the following properties:
\begin{eqnarray}
\Phi(x+ a) &= & e^{i \kappa a} \Phi(x),\\ \nonumber
\Phi(x+ N a) &=& \Phi(x),
\end{eqnarray}
where $a$ is the distance between two adjacent wells (lattice constant) and $\kappa= \lambda\ \frac{ 2\pi}{Na}$ is the pseudo-momentum with $\lambda= 0, \pm1, \pm2,...$. Using the Bloch theorem and the Linear Combination of Atomic Orbital (LCAO) method with no hybridization assumption, the lattice wave function of the $n^{\text{th}}$ energy level is $\Phi_{n}(x) = \sum\limits_{j=1}^{N} \ e^{ ika j} \chi^{(j)}_{n}(x)$ in which $\chi(x)$ is the solution of each atom.\\

The 3-fold symmetry potential that hinders the methyl group's rotation around the symmetry axis can be considered as $N=3$ periodic lattice with the lattice constant $a= \frac{2\pi}{3}$. Therefore, we take the Bloch periodic wavefunction and simply replace $x=\varphi$ to get 
\begin{eqnarray}
\Phi_{\lambda, n}(\varphi)= \sum\limits_{j=1}^{3}  e^{i \frac{2\pi }{3}  \lambda j }\chi^{(j)}_{n}(\varphi),
\end{eqnarray}
in which $\chi^{(j)}_{n}$ is the $n^{\text{th}}$ eigenfunction of the $j^{\text{th}}$ well (harmonic oscillator). The boundary condition, $\Phi(\varphi+ 2\pi)=\Phi(\varphi)$, results in $\lambda=0, \pm1$. Thus,
\begin{eqnarray}
\label{Eq_Torsional_states}
 \Phi_{\lambda=0,n}  &=& \frac{1}{\sqrt{3}}\ \left( \chi^{(1)}_{n} + \chi^{(2)}_{n} +  \chi^{(3)}_{n}\right),\\ \nonumber
  \Phi_{\lambda=1,n}  &=& \frac{1}{\sqrt{3}}\ \left( \chi^{(1)}_{n} + \epsilon^{*}\ \chi^{(2)}_{n} +  \epsilon \ \chi^{(3)}_{n}\right),\\ \nonumber
 \Phi_{\lambda=-1,n} &=&  \frac{1}{\sqrt{3}}\ \left( \chi^{(1)}_{n} + \epsilon\ \chi^{(2)}_{n} + \epsilon^{*}\ \chi^{(3)}_{n}\right),\\ \nonumber
\end{eqnarray}
 with $\epsilon= e^{i\frac{2 \pi}{3}}$. Because of the cyclic boundary condition, we have $\chi^{j}(\varphi+ \frac{2\pi}{3})= \chi^{j+1}(\varphi)$. Consequently, the $\lambda=0$ state has $A$ symmetry and the $\lambda=\pm1$ states have $E_{\pm}$ symmetry. We should also note that the above LCAO expansions are valid in the limit of no-hybridization assumption where the overlap between $\chi^{(j)}_{n}$ and $\chi^{(j')}_{n+1}$ is negligible. In that limit, the Hamiltonian is block diagonal and the corresponding Schr\"{o}dinger equation for the $n^{\text{th}}$ block is 
  \begin{eqnarray}
  H_{n}\  \Phi_{\lambda,n}& =& E_{\lambda,n}\  \Phi_{\lambda,n}\\ \nonumber
 \left( \begin{array}{ccc}
  \alpha _{n}& \beta_{n} & \beta _{n}\\
  \beta _{n}& \alpha_{n} & \beta _{n}\\
  \beta _{n}& \beta_{n} & \alpha _{n}
  \end{array} \right) \left( \begin{array}{c}
  1\\
  e^{\frac{2\pi i}{3} \lambda}\\
   e^{-\frac{2\pi i}{3} \lambda}
  \end{array}\right)&=& E_{\lambda,n} \left( \begin{array}{c}
   1\\
  e^{\frac{2\pi i}{3} \lambda}\\
   e^{-\frac{2\pi i}{3} \lambda}
  \end{array}\right)\\ \nonumber
  \Longrightarrow  E_{\lambda,n} &=& \alpha_{n} + 2 \beta_{n} \cos\left( \frac{2 \pi}{3} \lambda\right).
  \end{eqnarray} 
Here, $\alpha_{n}=E^{(0)}_{n}$ is the eigenenergy of each well and $\beta_{n}= - E^{(0)}_{n} \langle \Phi^{(j)}_{n}|\Phi^{(j+1)}_{n}\rangle$ is the overlap between the wavefunctions of the two adjacent wells. This overlap results in a splitting, $|\Delta E_{n}| = 3 \beta_{n}$, between the $A$ symmetry level and the $E_{\pm}$ symmetry levels. This is an analogy to the energy band gap in the solid state physics when we deal with a periodic lattice with large $N$. \\

The overlap between the wavefunctions of two wells provides a qualitative and yet informative description of the torsional eigenenergies. The symmetry of each well/harmonic oscillator's wavefunctions changes from the $n^{\text{th}}$ level to the $(n+1)^{\text{th}}$ level. Consequently, the sign of the overlap, $\beta_{n}$, or the splitting $\Delta E_{n}$ changes from the $n^{\text{th}}$ level to the $(n+1)^{\text{th}}$ level, and accordingly the ordering between the $A$ level and the $E_{\pm}$ levels changes. Moreover, in high energy levels, the wavefunction is less confined within a well and therefore its overlap with the neighbor's wavefunction is larger. This means, as we go higher and higher in energy, the splitting between $A$ symmetry and $E_{\pm}$ symmetry gets larger and larger and at some point the labeling $\{ A, E_{\pm}\}$ is not valid any more, because, the no-hybridization assumption breaks. Indeed, if the torsional energy is high enough, the barrier is not effective  any more and the methyl group behaves like a free rotor. This symmetry analysis is consistent with the numerical result presented previously.\\

  Now that we have some understanding of the spin symmetry, the torsional symmetry respectively, we proceed to the argument of the symmetry of the total wavefunction and the correlation between the spatial space and the spin space. One may take advantage of this correlation to create a noise protected qubit.

 \section{The symmetry of the Total Wavefunction}
 \label{Sec_Tot_Sym}
 We are interested in the solid phase of $X-CH_{3}$ molecule, where the translational and external rotation of this \textit{rigid} body or its center of mass is negligible. Excluding the spin degree of freedom, a methyl group has rotation, vibration and electronic degrees of freedom that are known as Rovibronic for short.  At low temperature (solid phase), the external rotational and translational motion are negligible and it is a fair assumption to consider a fixed molecular framework in which the nuclei rotate and/or vibrate near their equilibrium locations relative to a fixed origin (center of mass). Since electrons are much lighter than nuclei, an electron is much faster than a nucleus, and thus, its corresponding energy is very larger than the nuclear energy. This justifies the Born-Oppenheimer approximation \cite{BH54, BO27}, that is commonly used in the quantum chemistry to effectively separate the wave function into an electron part and a nuclear part, $\psi_{e} \otimes \psi_{n}$. In this approximation, one can treat the nuclei at a fixed geometry or in certain configurations in space with a slow motion and solve the Schr\"{o}dinger equation for the electron wavefunction only, yielding $\psi_{e}$ which is derived in the nuclear coordinates. In the second step, for each electron state, the nuclear wavefunction is obtained by including an effective potential in the Hamiltonian that serves as a replacement for each electron wavefunction. Indeed, in each electronic level, there is a set of eigenergies for the nuclear spins. \\
 
  Further, we need to consider the rotation-vibration part of the wavefunciton. In general, the rotation and the vibration are not two independent degrees of freedom, but, we might still be able to treat them separately. The vibrational energies are normally in the order of $1000\  cm^{-1}$, whereas the rotational energies are in the order of $10\  cm^{-1}$  \cite{BJ05}. Therefore, one can treat the rotational variables as a constant, and solve the Schr\"{o}dinger equation by considering the vibrational part only. In this method, similar to the Born-Oppenheimer approximation, for each vibrational state $\nu$, an effective potential $V_{\nu}$ is added to the rotational part of the Schr\"{o}dinger equation. \\
   
Considering the above approximations, the electronic, the vibrational, the rotational and the spin degrees of freedom are treated independently  and the total wave function is written as a product of them, 
\begin{equation}
\Psi_{\text{tot}}= \psi_{e} \otimes \psi_{\text{vib}} \otimes \psi_{\text{tor}} \otimes \psi_{\text{spin}} .
\end{equation}

  At low temperature, $\psi_{e}$ and $\psi_{\text{vib}}$ are mostly in their ground state which are symmetric functions \cite{BJ05}. Therefore, the symmetry of the total wavefunction is determined by the symmetry of the product of the internal rotation (torsional) eigenfunctions and the the spin eigenstates. \\
 
For fermionic  systems, the total wavefunction of three identical particles must be invariant under the $2 \pi/3$ rotation, because this rotation is equivalent to two particle exchanges: First, the $j^{\text{th}}$ particle with the $(j+1)^{\text{th}}$ one, and second the $(j+1)^{\text{th}}$ particle with the $(j-1)^{\text{th}}$ one, for all $j\in\{1,2,3\}$. 
This invariance under the rotation narrows the allowed combinations of the torsional eigenfunctions and the spin eigenstates to those that satisfy 
$$\text{Sym}(\psi_{\text{tor}}) \times \text{Sym}(\psi_{\text{spin}})= A,$$
 where $\text{Sym}(f)$ stands for the symmetry of $f$. Thus, the allowed combinations are $A \times A$, $E_{+} \times E_{-}$ and $E_{-} \times E_{+}$. This can be considered as a sort of \textit{symmetry correlation} between the spatial space and the spin space. One can take advantage of this symmetry induced coupling between the spin space and the torsional space to initialize the spin state of the methyl group in the noise protected subsystem.\\
 
Note that an implicit approximation is considered here. When a methyl groups is rotating in the presence of a hindering potential, the three protons are no longer perfectly indistinguishable to the environment, and hence, the spin Hamiltonian in Eq. \ref{Eq_Spin_Ham} and the corresponding spin eigenstates are an approximation. But as a first order approximation, we treat them as identical spins.

\section{Accessing the Noise Protected Subsystem of Methyl Groups}
\label{Sec_protected_state_Methyl}
The proposed experiment for encoding one logical qubit into the noiseless subsystem of a methyl group consists of two phases. First, we show that at cryogenic condition and at \textit{low field}, where the spin Zeeman splitting is much smaller than the torsional ground state energy splitting, the thermal equilibrium state is highly populated in the  $A \times A$ subspace. Second, we show that an interaction with a microwave field can, in principle, selectively \textit{convert} the population from the $A \times A$ subspace to the $E_{-} \times E_{+}$ subspace or to $E_{+} \times E_{-}$ subspace, depending on the applied being left or right circularly polarized. We elaborate on each of the above phases in the following sections.\\

\subsection{Long Lived State by Thermal Means}
\label{Sec_LLS_Thermal}
\textit{Long Lived States} (LLS) refers to those states that have an imbalance of population between different symmetry subspaces and have been experimentally studied in methyl groups by \cite{C71, IB12, M13}. LLS exhibit long relaxation times because they are polarized in terms of the symmetry label which is immune to the collective noise. LLS are considered as classical protected states because there is no coherence term between different symmetry subspaces. In the following, we review how an LLS in a methyl group can be implemented just by thermal means and discuss why this state is not capable of storing quantum information. We later use the LLS as an initial state that can be converted to a logical qubit. \\

We define \textit{symmetry polarized} states as a set of mixed states that are individually polarized in terms of the symmetry label, but are totally mixed in terms of the total spin magnetization label. They are denoted by $\rho_{A/E_{\pm}}$, and are given by
\begin{eqnarray}
\label{Eq_base_states}
\rho_{A} &:=& \frac{1}{4} \sum\limits_{m=-\frac{3}{2}}^{\frac{3}{2}} |A , m \rangle \langle A, m | = \frac{1}{4}\left(\begin{array}{c|c}
 \mathbb{1}_{4} & 0\\
 \hline
 0 & 0
\end{array} \right),\\\nonumber
\rho_{E_{+}} &:=& \frac{1}{2} \sum\limits_{m=-\frac{1}{2}}^{\frac{1}{2}} |E_{+} , m \rangle \langle E_{+}, m |=\frac{1}{2}\left(\begin{array}{c|c}
0 & 0\\
\hline
 0 & \begin{array}{c|c}
\mathbb{1}_{2} & 0\\
\hline
 0 & 0
\end{array}
\end{array} \right),\\ \nonumber
\rho_{E_{-}} &:=& \frac{1}{2} \sum\limits_{m=-\frac{1}{2}}^{\frac{1}{2}} |E_{-} , m \rangle \langle E_{-}, m |=\frac{1}{2}\left(\begin{array}{c|c}
0 & 0\\
\hline
 0 & \begin{array}{c|c}
0 & 0\\
\hline
 0 & \mathbb{1}_{2}
\end{array}
\end{array} \right).
\end{eqnarray}
An LLS is $\gamma-$polarized in terms of the symmetry label if there is an imbalance of population between the $A$ subspace and the $E$ subspace, and is given by
 \begin{eqnarray}
 \label{Eq_LLS}
 Q_{LLS}&= &\frac{(1+\gamma)}{2}\rho_{A} + \frac{(1-\gamma)}{2} \rho_{E}\\ \nonumber
 &=&\frac{ 1}{8}( \mathbb{1} +  \frac{ \gamma}{3}\ (\vec{\sigma}^{(1)}. \vec{\sigma}^{(2)} + \vec{\sigma}^{(2)}. \vec{\sigma}^{(3)}  + \vec{\sigma}^{(1)}. \vec{\sigma}^{(3)})),
 \end{eqnarray}
where $\rho_{E}=\frac{1}{2}\left(\rho_{E_{+}}+ \rho_{E_{-}}\right)$. Because of the scalar terms $\vec{\sigma}. \vec{\sigma}$ that are invariant under $\hat{\textbf{S}}_{\alpha}$, with $\alpha \in \{x,y,z\}$, the $Q_{LLS}$ is protected against all collective noise.\\ 

We show how to populate $Q_{LLS}$ by thermally cooling a methyl group. Consider an intermediate rigid rotor in the presence of a magnetic field with torsional eigenfunctions $\Phi_{\lambda, n}$ and spin eigenstates $|s, m\rangle$. As discussed in Section.\ref{Sec_Tot_Sym}, for a fermionic system, the product of $\Phi_{\text{tor}} \otimes \psi_{\text{spin}}$ must be symmetric under the $\frac{2 \pi}{3}$ rotation, which narrows the allowed combinations of the torsional eigenfunctions and the spin eigenstates to those with symmetry labels $\lambda \times s \in \{ A \times A, \ E_{\pm} \times E_{\mp}\}$. This \textit{symmetry correlation} between the spatial space and the spin space can be used for implementing a LLS.\\

\begin{figure}[h]
  \centering
\includegraphics[width=0.9\linewidth]{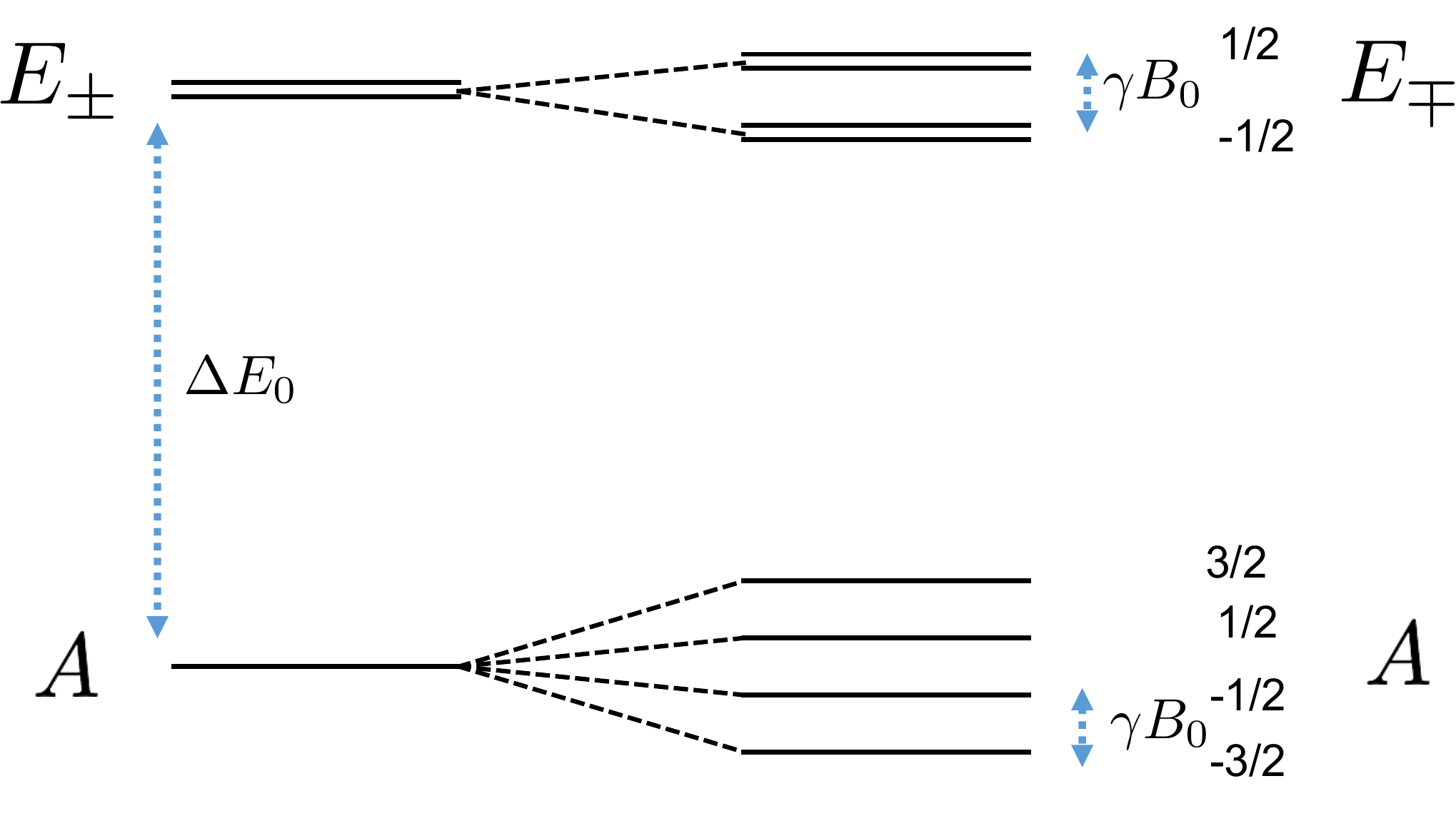}
  \caption[Eigenenergies of a Rigid Rotor with Medium Barrier Height]{  \label{fig_Eigen_Structure}Eigenenergies of a Rigid Rotor with Medium Barrier Height: Each torsional level with symmetry $\lambda$ is further split to more energy levels with spin symmetry $s$, such that the total symmetry is $\lambda \times s = A$. $\Delta E_{0}$ is the splitting between the $(\lambda= A, n=0)$ level and the $(\lambda=E_{\pm}, n=0)$ levels on the torsional space.}
\end{figure}

 At relatively low field, when the Zeeman splitting, $\gamma_{h} B_{0}$, is negligible compared to the torsional ground state splitting, $\Delta E_{0}$, the spatial Hamiltonian is dominant. Fig.\ref{fig_Eigen_Structure} schematically represents the energy diagram of the first ground state of a methyl group with a medium rotational barrier, in which $\Delta E_{0}$ is the splitting between the $(\lambda= A, n=0)$ level and the $(\lambda=E_{\pm}, n=0)$ levels. At relatively low temperature, when the system is cooled much below the torsional ground state splitting, $k_{B}T \ll \Delta E_{0} $, the $(\lambda =A, n=0)$ level in the torsional space is highly populated. If this temperature is still high compared to the Zeeman splitting, the $|s, m\rangle$ levels in the spin space are almost equally populated and the spin density matrix is almost equal to an identity in terms of the spin magnetization label (Eq.\ref{Eq_base_states}).\\

Thus, at low field and low temperature ($\gamma B_{0} \ll k_{b} T \ll \Delta E_{0}$) and according to the Boltzmann distribution, there is a significant population imbalance between the $\Phi_{A, n=0}$ and the $\Phi_{E_{\pm}, n=0}$ in the torsional space. Consequently, due to the spin-space symmetry correlation, there is an imbalance of population imbalance the $\rho_{A}$ and the $\rho_{E}$ in the spin Hilbert space. As a result, this thermal process corresponds to initializing the collective spin state of a methyl group in a $\gamma$-polarized long lived state, $Q_{LLS}$ with $\gamma = \tanh[ \Delta E_{0}/k_{b} T]$. This has been experimentally demonstrated in \cite{C71, IB12, M13}. \\ 

Note that the $Q_{LLS}$ that was introduced in Eq.\ref{Eq_LLS} is a classical mixture of $\rho_{A}$ and $\rho_{E}$, which are themselves mixed states. When neither the local spin Hamiltonian nor the interaction Hamiltonian distinguishes spin, one is not able to create coherence between the $A$ subspace and the $E$ subspace, and use $\{A/E\}$ as a logical basis for quantum information processing. In other words, when spins are indistinguishable, the spin Hamiltonian has a block diagonal form, and hence, the Hilbert space is a direct sum of the two subspaces, $\mathcal{H}= \mathcal{H}_{A} \oplus \mathcal{H}_{E}$. Nevertheless, the subspace $\mathcal{H}_{E}$ alone can be used for encoding a logical qubit, which is the subject of discussion in the next section.

\subsection{Protected State via the Electromagnetic Field Interaction}
\label{Sec_MW_irradiation}
As mentioned before, the $E$ subspace of three identical spins is decomposed into a product of two subsystems, $\mathcal{H}_{E}= \mathbb{C}^{2} \otimes \mathbb{C}^{2}$, where the first subsystem refers to the symmetry, $s\in \{E_{+}, E_{-}\}$ which can be used as a logical basis, and the second subsystem refers to the total magnetization, $m \in \{ \pm \frac{1}{2}\}$, which can be corrupted by noise, and hence, is not of interest. For storing quantum information, our target is to access the $E_{\pm}$ subsystem of the spin space individually without removing the degeneracy of these subspaces.  \\

One intuitive solution for accessing the $E_{+}/E{-}$ noiseless subsystem without breaking their degeneracy is to include a control Hamiltonian that selectively populates either the $ E_{+}$ subsystem (logical $|\bar{0}\rangle $) or the $ E_{-}$ subsystem (logical $|\bar{1}\rangle $). We propose to first prepare $Q_{LLS} $ with $\gamma \approx 1$, as it was explained in Section.\ref{Sec_LLS_Thermal}. This step is done at very low temperature and very low field, which initializes the collective spin state in $\rho_{A}$ that is polarized in terms of the spin symmetry label. In the next step, we apply a selective \enquote{$\pi$} pulse that converts the $\Phi_{A,0}$ to the $\Phi_{ E_{+},0}$ (or $\Phi_{ E_{-},0}$) on the torsional space. This leads to populating $\rho_{E_{-}}$ (or $\rho_{E_{+}}$) on the spin space without removing the degeneracy of them. We have not yet described how to implement these \enquote{$\pi$} pulses. This section is exploring the possibility of addressing the logical basis (or the $E_{\pm}$ subsystems) by using an interaction between the rotational degree of freedom of methyl groups and a circularly polarized external electromagnetic field.\\
 
We are inspired by microwave spectroscopy \cite{TS75, JAM03} which is a well known technique that uses the microwave irradiation to cause transitions between the rotational states of molecules in the gas phase. In this spectroscopy technique, the emission and the absorption of the electric dipole allowed transitions leads to extracting information about the geometry of rigid bodies such as the bond's length and angles, \cite{HB78, TS75}. We adopt this technique and apply it to our case, where at low temperature the system is in the solid phase rather than the gas phase. The barrier height can be different at solid phase versus the gas phase but not by orders of magnitude.  At low field, we ignore the spin space and just focus on the spatial space transitions. Consider a rigid rotor in a symmetric top molecule, $X-CH_{3}$, whose torsional ground state splitting $\Delta E_{0}$ is in the range of GHz. We explore whether a circularly (right or left) polarized microwave filed that is on resonance with $\Delta E_{0}$, induces a transition between the $\Phi_{A,0}$ and the $\Phi_{ E_{\pm},0}$ or not.\\

A molecule with a permanent dipole moment $\vec{\textbf{d}}$ interacts with a time varying electromagnetic field, $\vec{\textbf{E}}(\textbf{r}, t)$, via an electric dipole Hamiltonian, $H_{d} =-\hbar\  \vec{\textbf{d}}. \vec{\textbf{E}}(\textbf{r},t)$. For a point charge, the dipole moment is $\vec{\textbf{d}}(\vec{\textbf{r}}) = q\ \vec{\textbf{r}}$, and for a charge distribution, the dipole moment is 
$$\vec{\textbf{d}}(\vec{\textbf{r}}) = \int\ \rho_{e}(\vec{\textbf{r}}-\vec{\textbf{r}'}) \ \vec{\textbf{r}'}\  d^3\vec{\textbf{r}'},$$
where $ \rho_{e}(\vec{\textbf{r}})= |\psi_{e}(\vec{\textbf{r}})|^2$ is the electron charge distribution. If the electromagnetic wavelength is much larger than the molecule size,  the field is approximately constant across the molecule, $\vec{\textbf{E}}(\textbf{r}, t)\sim\vec{\textbf{E}}(t)$. According to Fermi's Golden Rule, the probability of transition from the $i^{\text{th}}$ level to the $j^{\text{th}}$ level due to an interaction Hamiltonian $H_{\text{int}}=H_{d}$ is given by
\begin{eqnarray}
W_{i \rightarrow j} &=& \frac{2 \pi}{\hbar } \ |\langle \psi_{i} | H_{\text{int}} | \psi_{j} \rangle|^{2}, \\ \nonumber
&=& 2\pi \ \left( \int \psi^{*}_{i} ( \vec{\textbf{r}})\  \rho_{e}(\vec{\textbf{r}}) \ \vec{\textbf{r}}.\vec{\textbf{E}}(t) \ \psi_{j} (\vec{\textbf{r})}d^{3}\vec{\textbf{r}}\right)^{2}.
\end{eqnarray}
 The transition rate is proportional to an integral which is non-zero (allowed transitions) if and only if the integrand is a totally symmetric function. Thus, a transition from $\psi_{i}$ to $\psi_{j}$ is dipole allowed iff $\text{Sym}(\psi_{i}) \times\text{Sym}(H_{\text{d}}) \times \text{Sym}(\psi_{j}) =A$.\\

Excluding the spin space and the external rotations, $\psi_{i}( \vec{\textbf{r}})$ is the internal rotation-vibration wave function. At low temperature, the vibrational wavefunction is symmetric, and therefore, the symmetry of $\psi_{i}( \vec{\textbf{r}})$ is determined by the symmetry of the torsional wavefunction, $\Phi_{\lambda, n}(\varphi)$. Moreover, the electron charge distribution $ \rho_{e}(\vec{\textbf{r}})= |\psi_{e}(\vec{\textbf{r}})|^2$, is also symmetric. Therefore, to find the selection rule (forbidden vs allowed transitions), it is sufficient to analyze the symmetry of $\Phi_{\lambda,n}\ \vec{\textbf{r}}.\vec{\textbf{E}}(t)  \ \Phi_{\lambda', n'}$. Since we are only interested in the allowed transitions between $\Phi_{A,0}$ and $\Phi_{E_{\pm},0}$, the only unknown component in the above integral, is the symmetry of $\vec{\textbf{r}}.\vec{\textbf{E}} (t)$. \\

A circularly polarized electromagnetic field is $\vec{\textbf{E}}_{\pm}(t) = \mathcal{E}_{0}\  (\hat{x} \pm i \hat{y})\  e^{ i \omega t}$, where $\mathcal{E}_{0}$ is the amplitude of the field \cite{J62}. Thus,
\begin{equation}
 \vec{\textbf{r}}.\vec{\textbf{E}}_{\pm} (t) = \mathcal{E}_{0}\  (X \pm i Y) \ e^{i \omega t}. 
\end{equation}

According to the group theory, one can find the symmetry of any spatial function by applying the \textit{symmetry projection} operators to it \cite{Vv01}. In particular, in case of $\textbf{C}_{3}$ group, we apply $\hat{\textit{P}}_{A}, \hat{\textit{P}}_{E_{+}}$ and $\hat{\textit{P}}_{E_{-}}$ on $(X + i Y)$ and obtain
\begin{eqnarray}
\label{Eq_symm_MW}
\hat{\textit{P}}_{A}.(X + i Y) &=& 0,  \\ \nonumber 
\hat{\textit{P}}_{E_{+}}.(X + i Y) &=& 0, \\ \nonumber 
\hat{\textit{P}}_{E_{-}}.(X + i Y) &=& (X + i Y),
\end{eqnarray}
where 
\begin{eqnarray}
\hat{\textit{P}}_{A} &=& \frac{1}{3} \left(  \mathbb{1} + R_{+} + R_{-}\right), \\ \nonumber
\hat{\textit{P}}_{E_{+}} &=& \frac{1}{3} \left(  \mathbb{1} + \epsilon \ R_{+} + \epsilon^{*} \ R_{-}\right), \\ \nonumber
\hat{\textit{P}}_{E_{-}} &=& \frac{1}{3} \left(  \mathbb{1} + \epsilon^{*} \ R_{+} + \epsilon \  R_{-}\right).\\ \nonumber
\end{eqnarray}
Here, $R_{\pm}$ are $\pm \frac{2\pi}{3}$ rotation around the symmetry axis and $ \mathbb{1}$ is the \textit{no-rotation} operator. Eq.\ref{Eq_symm_MW} means that the electric dipole interaction with a \textit{right circularly polarized} electromagnetic field has $E_{-}$ symmetry. Therefore, for this particular interaction, the transition from $\Phi_{A,0}$ to $\Phi_{E_{+},0}$ is symmetrically allowed. Similarly, a \textit{left circularly polarized} field has the $E_{+}$ symmetry that induce transition from $\Phi_{A,0}$ to $\Phi_{E_{-},0}$. This observation may enable us to selectively populate the $E_{+}$ or the $E_{-}$ subspaces on the spin space as a result of the correlation between the torsional space and the spin space. \\

We introduce an effective Hamiltonian that takes into account both the electric dipole allowed transition due to the interaction with the right circularly polarized $\mu w$, and the spin-space correlation due to Pauli exclusion principle,
\begin{eqnarray}
\label{Eq_effective_ham}
H_{\text{eff}}=\kappa \ \left( |\Phi_{E_{-},0}\rangle \langle \Phi_ {A,0}|  \otimes \sum\limits_{m\in\{\pm 1/2\} } | E_{+},m\rangle  \langle A,m|\right) + h.c. .
\end{eqnarray}
The strength of the coupling, $\kappa$, depends on the field amplitude $\mathcal{E}_{0}$ and the component of the electric dipole operator $\vec{\mathbf{d}}$ that is perpendicular to the symmetry axis. Before turning on the $\mu w$, and after the cooling, our system is initialized in
\begin{equation}
\rho_{0} = |\Phi_{A,0}\rangle\langle \Phi_{A,0}| \otimes \sum\limits_{m=-3/2}^{3/2} \ \frac{1}{4} \ | A,m\rangle \langle A, m|.
\end{equation}
Once we turn on a right circularly polarized $\mu w$ for a time $\tau_{0}$ so that $2\pi \kappa \tau_{0} = \pi$, the collective spin system evolves to
\begin{eqnarray}
 \rho_{\text{spin}}&=& Tr_{\text{tor}}\left[ e^{-i H_{\text{eff}} \tau_{0}} \ \rho_{0}\ e^{+i H_{\text{eff}} \tau_{0}} \right]\\ \nonumber
&=& \frac{1}{4} \left[ \sum\limits_{m\in\{\pm 3/2\} }  |A,m\rangle \langle A,m| + \sum\limits_{m\in\{\pm 1/2\} }  \ |E_{+},m \rangle \langle E_{+},m| \right].
\end{eqnarray}
 Therefore, the cooling followed by the $\mu w$ irradiation effectively converts the population from $|A,\pm \frac{1}{2}\rangle$ to $|E_{+},\pm \frac{1}{2}\rangle$ on the spin space. But, because of the mismatch of the dimension of the two subspaces, there is some undesired population left in the $|A,\pm \frac{3}{2}\rangle$ levels. Nevertheless, this is not an issue because the scalar coupling between the methyl group and an external spin, such as carbon, shifts the frequency of $m=\pm1/2$ from that of $m=\pm 3/2$ and makes them distinguishable. Therefore, one can, in principle, post-select the $E_{+}$ events at the expense of decreasing the probability of success,
\begin{equation}
\rho^{\text{post-select}}_{\text{spin}}=\rho_{E_{+}}= \frac{1}{2}\ \sum\limits_{m\in\{\pm 1/2\} }  \ |E_{+},m \rangle \langle E_{+} ,m|.
\end{equation}
Similarly, the system is initialized in $\rho_{E_{-}}=\frac{1}{2}\ \sum\limits_{m\in\{\pm 1/2\} }  \ |E_{-},m \rangle \langle E_{-} ,m| $ via an interaction with the left circularly polarized $\mu w$. Thus, we have access to the logical basis states, $|\bar{0}\rangle\langle \bar{0}| $ and $|\bar{1}\rangle\langle \bar{1}| $, individually. By the right choice of the intensity of the left and right circularly polarized fields, one can create 
\begin{eqnarray}
Q_{\text{logic}}:=\frac{(1+\beta)}{2}\rho_{E_{+}} +\frac{(1-\beta)}{2}\rho_{E_{-}}.
\end{eqnarray}
$Q_{\text{logic}}$ is still a classical mixture of the logical states. To prepare any arbitrary superposition of the logical states, we require to implement an \enquote{$X$} gate which takes $|\bar{0}\rangle \leftrightarrow|\bar{1}\rangle $ and a \enquote{$Z$} which takes $|\bar{0}\rangle \rightarrow |\bar{0}\rangle $ and $|\bar{1}\rangle \rightarrow - |\bar{1}\rangle$. Given the logical gates, \enquote{$X$} and \enquote{$Z$}, one has universal control over a single qubit \cite{NCu00}. In the following, we now show implementing these logical gates is feasible.\\

Suppose we are able to implement the following rotations:
\begin{eqnarray}
&&U_{\pm}(\theta, \phi):  |\Phi_{A,0}\rangle \otimes  \ |A,m \rangle \\ \nonumber
&&\longrightarrow   \cos\frac{\theta}{2}\  |\Phi_{A,0}\rangle \otimes  \ |A,m \rangle  + e^{i \phi}
 \sin\frac{\theta}{2}\  |\Phi_{E_{\mp},0}\rangle \otimes  \ |E_{\pm}, m \rangle, 
\end{eqnarray}
 for $m= \pm \frac{1}{2}$. Indeed, $U_{\pm}$ is an arbitrary rotation from the ground state to the first excited state, where both the torsional and the spin degrees of freedom are considered. Given $U_{\pm}(\theta, \phi)$, the control gates is implemented by the following sequences
\begin{eqnarray}
X&:=& U_{-}( \pi, 0) \ U_{+}(\pi, 0) \ U_{-}( \pi, 0), \\ \nonumber
Z&:=& U_{-}(2 \pi, 0)\  U_{+}(2 \pi, 0)\  U_{-}(2 \pi, 0),
\end{eqnarray} 
 which act as
 \begin{eqnarray}\nonumber
X: && \frac{1}{2}\  \left[ |\Phi_{E_{\mp},0}\rangle \langle \Phi_{E_{\mp},0}| \otimes \sum\limits_{m\in\{\pm 1/2\} }  \ |E_{\pm},m \rangle \langle E_{\pm} ,m| \right]\\ \nonumber
&\longrightarrow & \frac{1}{2} \left[ |\Phi_{E_{\pm},0}\rangle \langle \Phi_{E_{\pm},0}| \otimes \sum\limits_{m\in\{\pm 1/2\} }  \ |E_{\mp},m \rangle \langle E_{\mp} ,m| \right]\\ \\\nonumber \\  \nonumber
Z: && \frac{1}{2}\  \left[ |\Phi_{E_{\mp},0}\rangle \langle \Phi_{E_{\mp},0}| \otimes \sum\limits_{m\in\{\pm 1/2\} }  \ |E_{\pm},m \rangle \langle E_{\pm} ,m| \right]\\ \nonumber
&\longrightarrow \hskip 0.2 cm   \pm & \frac{1}{2}\  \left[ |\Phi_{E_{\mp},0}\rangle \langle\nonumber \Phi_{E_{\mp},0}| \otimes \sum\limits_{m\in\{\pm 1/2\} }  \ |E_{\pm},m \rangle \langle E_{\pm} ,m| \right]\\ \nonumber
 \end{eqnarray}
The above operators satisfy the commutation relations of Pauli operators, e.g., $[X, Z] = 2 i Y$ and they obey the anti-commutation rule. Therefore, the set of operators generated by the multiplications of these logical gates and their Hermitian conjugation construct the algebra of a qubit. For completeness of the discussion, in order to implement any arbitrary coherent superposition of the logical basis, i.e., $|\psi\rangle= \cos\frac{\alpha}{2} \ |\bar{0}\rangle +e^{i\beta}  \sin \frac{\alpha}{2}\  |\bar{1}\rangle$, one needs to implement the following rotation
\begin{eqnarray}\nonumber
&&R(\alpha, \beta) :=U_{-}( -\pi, \pi) \ U_{+}(\alpha, -\beta) \ U_{-}( \pi, \pi),\\ \nonumber
&\text{which acts as}&\\ \nonumber
&&R(\alpha, \beta)|\Phi_{E_{-},0}\rangle \otimes |E_{+} , m\rangle\\ \nonumber
&&  \rightarrow \cos\frac{\alpha}{2} |\Phi_{E_{-},0}\rangle \otimes |E_{+}, m\rangle
+ e^{i\beta}\sin\frac{\alpha}{2} \  |\Phi_{E_{-},0}\rangle\nonumber
\end{eqnarray}
This rotation can be achieved by controlling the intensity and the phase of the circularly polarized $\mu w$ fields. \\

In summary, we propose to prepare a logical qubit in a methyl group encoded in spin and rotational degree of freedom. To choose the right molecule, we desire a symmetric top molecule, $X-CH_{3}$, where the rotational barrier of the methyl group is in the intermediate regime, i.e., it is neither a free rotor nor a firm rotor. We also desire that the energy spliting between the $(n=0, \lambda =A)$ level and  the $(n=0, \lambda =E_{\pm})$ of the torsioanl space to be in the range of GHz. By thermal equilibration at cryogenec temperature (about 0.3 K if $ \Delta E_{0} \approx 6.5$ GHz), the  $Q_{LLS}$ is well populated. This step is at nearly zero magnetic field but if one desire to confirm that the LLS has been prepared, then, one should rapidly warm up the system and transport it into a large magnetic field and measure the NMR spectrum. According to the previous studies \cite{IB12, M13}, one would expect to observe two peaks on the proton channel that have equal intensity and are anti-phase and observe four peaks on the carbon channel that have unequal intensities but they are two by two anti-phase. To convert this LLS into the logical basis subspace, one can apply a circularly polarized microwave electric field that is tuned to be on resonance with the torsional ground state splitting. Given the dipole moment of the molecule, we control the intensity, the phase and the timing of the microwave field so that we implement an effective $\pi$ rotation between $\Phi_{A,0}$ and $\Phi_{ E_{+},0}$ or $\Phi_{E_{-}, 0}$ torsional states. To confirm that the logical states have been prepared, one can measure the NMR spectra once again by rapidly transporting to room temperature and high magnetic field. We expect to observe the same anti-phase peaks on the proton channel but on the carbon channel we should see the signature of an initially prepared logical qubit. Particularly, the two peaks that are associated with $m=\pm \frac{3}{2}$ on the carbon channel remain the same as that of the LLS but the other two peaks that are associated with $m=\pm \frac{1}{2}$ will not be anti-phase anymore. This indicates that the spin population is converted from the $|A, m=\pm\frac{1}{2}\rangle$ states to the $|E_{\pm}, m=\pm\frac{1}{2}\rangle$ states. The absolute value of the phase indicates whether we prepared the logical $|\bar{0}\rangle$ ($E_{+}$ state) or the logical $|\bar{1}\rangle$ ($E_{-}$ state). In principle, any coherent superposition state, $|\psi\rangle= \cos\alpha \ |\bar{0}\rangle + \sin \alpha \ |\bar{1}\rangle$, can be implemented by controlling the intensity and the phase of the microwave electric field.

\section{Conclusion}
The presented work seeks for novel means of storing quantum information in the noise protected subspace of a group of identical spins. In particular, we investigated the symmetry of the spin degree of freedom and the rotational degree of freedom of methyl groups respectively and discussed the symmetry correlation between the two subspaces. That correlation has been used for creating classical states with long relaxation time, and in this report, we moved a step further to propose a scheme that uses this correlation for creating a logical qubit that is robust against collective noise. Our analysis provides a proof-of-principle experiment that relies on symmetry allowed transitions due to the interaction between the dipole moment of the molecule with a circularly polarized microwave field.\\ 
We acknowledge support from the Canada First Research Excellence Fund (CFREF), Industry Canada, the Natural Sciences and Engineering Research Council of Canada (NSERC RGPIN-418579), Canada Excellence Research Chairs Program (CERC 215284), the Canadian Institute for Advanced Research, the Quantum NanoFab and the Province of Ontario.
\
\bibliography{Bibliography.bib}


\end{document}